\documentclass[aps,showpacs,preprintnumbers,amsmath,amssymb,nofootinbib,showkeys,preprint]{revtex4}

\usepackage{graphicx}
\usepackage{color}

\def \be  {\begin{equation}}
\def \ee  {\end{equation}}
\def \ee  {\end{equation}}
\def \bea {\begin{eqnarray}}
\def \eea {\end{eqnarray}}

\newcommand{\nn}{\nonumber}

\begin{document}

\preprint{ECTP-2013-03}

\title{Measurable Maximal Energy and Minimal Time Interval}

\author{Eiman ABOU EL DAHAB}
%\email{eaboeldahab@gmail.com}
\affiliation{Faculty of Computer Science, MTI University, 11571 Cairo, Egypt}

\author{Abdel Nasser~TAWFIK\footnote{http://atawfik.net/}}
%\email{a.tawfik@eng.mti.edu.eg}
\email{atawfik@cern.ch}
\affiliation{Egyptian Center for Theoretical Physics (ECTP), MTI University, 11571 Cairo, Egypt}

\date{\today}

\begin{abstract}
The possibility of finding the measurable maximal energy and the minimal time interval is discussed in different quantum aspects. It is found that the linear generalized uncertainty principle (GUP) approach gives a non-physical result. Based on large scale Schwarzshild solution, the quadratic GUP approach is utilized. The calculations are performed at the shortest distance, at which the general relativity is assumed to be a good approximation for the quantum gravity and at larger distances, as well. It is found that both maximal energy and minimal time have the order of the Planck time. Then, the uncertainties in both quantities are accordingly bounded.  Some physical insights are addressed. Also, the implications on the physics of early Universe and on quantized mass are outlined. The results are related to the existence of  finite cosmological constant and minimum mass (mass quanta).

\end{abstract}. 


\maketitle

%%%%%%%%%%%%%%%%%%%%%%%%%%%%%%%%%%%%%%%%%%%%%%%%%%%%%%%%%%%%%%%%%%%%%%
%%%   Section I
%%%%%%%%%%%%%%%%%%%%%%%%%%%%%%%%%%%%%%%%%%%%%%%%%%%%%%%%%%%%%%%%%%%%%%

\section{Introduction}
\label{sec:intr}

About fifty years ago, Shapiro pointed out that the possible time delay resulting from the observation that light appearing to slow down as it passes through a gravitational potential could be measured within our solar system \cite{td1,td2,Tawfik:2012hz}. A proposal of existing a minimal measurable time interval dates back to several decades. Furthermore, utilizing the fundamental limits governing mass and size of any physical system, Salecker and Wigner \cite{wigner57,wigner58} suggested that a minimum time interval can be even registered. They proposed the use of a quantum clock in measuring distances between events in spacetime. Although, the events are supposed to be macroscopic, measuring rods are avoided \cite{wigner58}. This quantum clock is given as constrains on smallest accuracy and maximum running time as functions of mass and position uncertainties. The Wigner's second constrain is assumed to be more severe than the Heisenberg uncertainty principle. The latter requires that only one single simultaneous measurement of both energy and time can be accurate. Wigner's contains assume that repeated measurements should not disturb the clock. On the other hand, the clock itself should be able to register time over its total running period, accurately.  

Taking into account quantum mechanical and general relativistic effects, Amelino-Camella derived some limitations on the measurability of spacetime distances \cite{disct3}. The detectability of quantum spacetime foam with gravitational wave interferometers has been addressed in Ref. \cite{disct1}, where the authors criticized the measurability limits of the smallest quantum distances. Operative definitions for quantum distances and elimination of the contributions from total quantum uncertainty are suggested \cite{disct2}. 

Barrow applied Wigner's inequalities to describe the quantum contains on black holes \cite{barrow96}. It is found that the black hole's running time should be correspondent to the lifetime of Hawking radiation. The latter is calculated under the assumption that the black hole behaves as a black body. Therefore, the Stefan-Boltzmann law can be utilized. Also, it is found that the Schwarzshild radius $r_s$ is assumed to be correspondent to Wigner's size constrain. Furthermore, the information processing power of a black hole is estimated by the emitted Hawking radiation. 
Based on generalized uncertainty principle (GUP), the resulting lifetime difference depends on black hole relative mass and the difference between black hole mass with and without GUP is not negligible \cite{Tawfik:2013uza}.

Recently, another fundamental limit was suggested. The existence of a minimal length is supposed to be one of the most interesting predictions of some approaches related to quantum gravity (QG) such as string theory (ST) \cite{st1a,st1b}.  Accordingly, the strings are conjectured not to interact at distances smaller than their size.  In cosmological aspects, it has been shown that the horizon is not defined at scales smaller than the Planck scale \cite{BHGUP1}.  This leads to generalizing Heisenberg uncertainty principle \cite{guppapers}. The insight of such generalization is taking into account the trace of gravity in the Heisenberg uncertainty principle. At Planck energy scale, the corresponding $r_s$ becomes comparable to the Compton wavelength. The two quantities become approximately equal to the Planck length.  These observations combined with {\it gedanken} experiments and rigorous derivations suggest that GUP is limited to some scales \cite{guppapers,BHGUP1,BHGUP2a,BHGUP2b,BHGUP3,BHGUP4,BHGUP5,BHGUP6,kmm,kempf,brau,Hossenfelder:2012jw}, 
\bea 
\Delta x\; \Delta p & \geq & \frac{\hbar}{2} \left[1 - 2 \alpha \langle p\rangle + 4 \, \alpha^2\, \langle p^2 \rangle  \right],   \label{uncert1}
\eea
where $p^2 = \sum \limits_{j} p_{j} p_{j}$ and $\alpha = \alpha_0/M_{pl}c= \alpha_0 \ell_{pl}/\hbar$. $M_{pl}$ and $M_{pl} c^2$ being Planck mass and energy, respectively. It would be assumed that $\alpha_0$, which is a dimensionless number, is not far from unity.
It was shown that the inequality given in Eq. (\ref{uncert1}) is equivalent to the modified Heisenberg algebra \cite{kmm}
\bea [x_i,\, p_j] = i\, \hbar\, \left[ \delta_{i j} - \alpha\left(p\, \delta_{i j} + \frac{p_i\, p_j}{p}\right) + \alpha^2 \left(p^2\, \delta_{i j} + 3 p_i \, p_k\right)\right], \label{com1} 
\eea
which in turn ensures, via the Jacobi identity, that $[p_i,\, p_j]=0$ \cite{kempf}. Eq. (8) of Ref. \cite{kempf} gives the commutation relation $[x_i,\, x_j]$. In a series of papers, the effects of GUP on quark gluon plasma \cite{Elmashad:2012mq}, compact stars \cite{Ali:2013ii}, inflationary era of the universe \cite{Tawfik:2012he}, Lorentz invariance violation \cite{Tawfik:2012hz} and the possible modification of Newton's law of gravitation \cite{Ali:2013ma} have been investigated.

Itzhaki considered the uncertainty principle and utilized the Schwarzshild solution in large scale in order to estimate the minimal measurable time interval \cite{itzhaki}. He found that the uncertainty in time measurement depends on the distance separating the observer from the event, the clock's accuracy and size, and the time taken by photon to reach the observer. Assuming distances, in which General Relativity (GR) offers a good approach for QG, then the shortest distance $x_c=\beta\, (G \hbar /c^3)^{1/2}$, where $\beta$ is an arbitrary parameter. The minimum error in the time measurement is estimated as
\bea
\Delta\, t &=& \sqrt{\frac{8 \, G\, \hbar}{c^5}\, \ln\left(\frac{x}{x_c}\right)}=2\,  \sqrt{2\, \ln\left(\frac{x}{x_c}\right)}\; t_{pl}, \label{eq:dlt}
\eea
where $t_{pl}=\sqrt{G\, \hbar/c^5}$ is the Planck time. This expression, Eq. (\ref{eq:dlt}), is valid at distance \hbox{$x>x_c\, \exp(2/\beta^2)$}. The shortest distance defines the scale up to which GR remains a good approximation to QG. The corresponding minimum error in the energy is given by
\bea
\Delta\, E &=& \sqrt{\frac{\hbar \, c^5}{2\, G \, \ln\left(\frac{x}{x_c}\right)}} = \sqrt{\frac{1}{2 \,  \ln\left(\frac{x}{x_c}\right)}}\;  \frac{\hbar}{t_{pl}}.
\eea
Then, minimal time and maximal energy at $x_c<x<x_c\, \exp(2/\beta^2)$, respectively, read
\bea
\Delta \, t_{min} &=& \frac{x_c}{c}\left[\frac{2}{\beta^2}+\ln\left(\frac{x}{x_c}\right) \right],\\
\Delta \, E_{max} &=& \frac{c^4}{2 \, G}\, x_c = \frac{\hbar \, x_c}{2 \, c}\; \frac{1}{t_{pl}^2}. 
\eea

\section{Maximal Energy and Minimal Time from Generalized Uncertainty Principle}
\label{sec:minET}

As introduced in section \ref{sec:intr}, a minimal measurable time and a maximal measurable energy can be deduced from  linear \cite{guppapers,linearr2,BHGUP1,BHGUP2a,BHGUP2b} and quadratic \cite{quardt1} GUP approaches.

\subsection{Linear GUP approach} 

In linear generalized uncertainty principle \cite{guppapers,linearr2,BHGUP1,BHGUP2a,BHGUP2b}, the uncertainty in time  reads
\bea
\Delta t & \geq & \frac{1}{2}\, \frac{\hbar}{\Delta E} \left[1 - 2\, \frac{\alpha}{c}\, \Delta E \right]  = \frac{\hbar}{2\, \Delta E} - \frac{\alpha}{c}\, \hbar,
\eea
implying that the physical limits require $2 \alpha \Delta E < c$. The minimum measurable time interval $\Delta t_m$ is to be deduced under the condition that the derivative $d\, \Delta \, t / d\, \Delta E$ vanishes. Then, 
\bea
- \frac{\hbar}{2\, (\Delta E)^2} &=& 0,
\eea
which leads to
\bea
\Delta\, E_{max} &=& \infty, \\
\Delta\, t_{min} &=& -\frac{\alpha}{c} \, \hbar= -\frac{\alpha_0}{M_{pl}\, c^2} \, \hbar,
\eea
where $\alpha$ is replaced by $\alpha_0/M_{pl} c$. The measurable maximal energy gets infinite while the measurable minimal time interval has a negative value. Both results are obviously non-physical. While $\Delta\, E$ violates the conservation of energy, $\Delta\, t$ shows that the direction of the arrow of time becomes opposite.

\subsection{Quadratic GUP approach } 

Applying the quadratic generalized uncertainty principle \cite{quardt1}, the uncertainty in time taken by photon to reach the observer is given as
\bea
\Delta t & \geq & \frac{1}{2}\, \frac{\hbar}{\Delta E} \left[1 - 2\, \frac{\alpha}{c}\, \Delta E + 4 \, \left(\frac{\alpha}{c}\right)^2 \, (\Delta \, E)^2 \right]  = \frac{\hbar}{2\, \Delta E} - \frac{\alpha}{c}\, \hbar + 2\, \hbar\,  \left(\frac{\alpha}{c}\right)^2 \, \Delta \, E.
\eea 
Therefore, the minimum measurable time interval $\Delta t_m$  occurs at
\bea
- \frac{\hbar}{2\, (\Delta E)^2} + 2 \,  \left(\frac{\alpha}{c}\right)^2 \, \hbar &=& 0,
\eea
leading to
\bea
\Delta\, E_{max} &=& \frac{c}{2\, \alpha}=\frac{E_{pl}}{2\, \alpha_0},\\
\Delta\, t_{min} &=& \frac{\alpha}{c}\, \hbar = \frac{\alpha_0}{M_{pl}\, c^2}\, \hbar,
\eea
where $E_{pl}=M_{pl}\, c^2$. We notice that the uncertainty in time is the same as the one obtained in the linear GUP approach but with a positive sign. Furthermore, the uncertainty in energy becomes finite. It is directly related to the Planck energy.

\subsection{Comparison between linear and quadratic GUP approaches}

Tab. \ref{table1b} summarizes a short comparison between quadratic \cite{guppapers,BHGUP1,BHGUP2a,BHGUP2b,BHGUP3,BHGUP4,BHGUP5,BHGUP6,kmm,kempf} and linear \cite{Ali:2010yn,Das:2010zf} GUP approaches in different aspects, namely Heisenberg algebra, minimal length uncertainty and maximal moment uncertainty and maximal measured moment. An extensive comparison is given in Ref. \cite{TD2013}. The parameter $\beta$ is related to $\alpha$, namely $\beta=\alpha^2$. For the linear approach, it is assumed that the higher orders of $\alpha$ vanish.

\begin{table}[htb]
\begin{center}
\begin{tabular}{| l || l | l |}
\hline \hline
  & Quadratic GUP \cite{guppapers,BHGUP1,BHGUP2a,BHGUP2b,BHGUP3,BHGUP4,BHGUP5,BHGUP6,kmm,kempf} & Linear GUP  \cite{Ali:2010yn,Das:2010zf} \\ 
\hline  \hline
Heisenberg Algebra & $\left[x,p\right]= i \hbar \left(1+\beta p^{2}\right)$ & $\left[x,p\right]= i \hbar \left(1-\alpha p +2 \alpha ^{2} p^{2} \right)$\\ 
\hline 
Minimal length uncertainty $\Delta x$ & $ \hbar \sqrt{\beta} $ & $ \hbar \alpha $    \\ 
\hline 
Maximal moment uncertainty $\Delta p$ & Undetermined & $M_{pl} c/\alpha _{0} $ \\ 
\hline 
Maximal moment $P_{max} $ & Divergence  & $1/(4\alpha)$   \\  
\hline 
& & String Theory  \\ 
Corresponding Theories & String Theory & Doubly Special Relativity \\ 
& & Black Hole Physics  \\
\hline 
\hline 
\end{tabular}
\caption{A comparison between quadratic \cite{guppapers,BHGUP1,BHGUP2a,BHGUP2b,BHGUP3,BHGUP4,BHGUP5,BHGUP6,kmm,kempf}  and linear \cite{Ali:2010yn,Das:2010zf}  GUP approaches in Heisenberg algebra, minimal length uncertainty and maximal moment uncertainty and maximal measured moment \cite{TD2013}. $\beta$ ($\alpha$) being quadratic (linear) GUP parameter.  }\label{table1b}
\end{center}
\end{table}

\section{Itzhaki Model and Generalized Uncertainty Principle}
\label{sec:Itzhaki}

In this section, we estimate the minimal measurable time and the maximal measurable energy from the Schwarzshild solution in large scale using the quadratic GUP approach. Comparing to the linear GUP approach, the quadratic one assures physical results. We implement Itzhaki model taking into consideration the quadratic GUP approach.  

\subsection{At the shortest distance  $x_c$}

The uncertainty in time as estimated from Schwarzshild solution down to a distance $x_c$ which define the scale up to which GR remains a good approximation to QG \cite{itzhaki} reads
\bea
\Delta\, t & \geq & \frac{\hbar}{2\, \Delta E} + G \frac{\Delta E}{c^5}. 
\eea
Then, the maximal measurable energy and minimal measurable time interval are given as 
\bea
\Delta\, E_{max} &=& c^2\, \sqrt{\frac{\hbar \, c}{2 \, G}}=\frac{\hbar}{\sqrt{2}}\;\frac{1}{t_{pl}}, \\
\Delta\, t_{min} &=& \frac{1}{c^2}\, \sqrt{\frac{2\, G \, \hbar}{c}}=\sqrt{2}\; t_{pl}.
\eea 
It is obvious that both quantities are positive and depend on the Schwarzshild radius which is related to the black hole mass, $r_s=(2 G/c^2)\, m$. It is worthwhile to note that both quantities are related to the Planck time $t_{pl}$ and accordingly, they are bounded.

When applying the quadratic GUP approach and when the time taken by photon to reach the observer  is taken into consideration, then the total uncertainty in time reads  
\bea
\Delta \, t_{total} & \geq & \frac{\hbar}{2\, \Delta E} - \frac{\alpha}{c}\, \hbar + 2\, G \frac{\Delta \, E}{c^5} \ln\left(\frac{x}{x_c}\right) .
\eea
Then, the maximal measurable energy and minimal measurable time interval, respectively, are
\bea
\Delta \, E_{max} &=& \frac{1}{2}\, \sqrt{\frac{c^5\, \hbar}{G\, \ln\left(\frac{x}{x_c}\right) }} =  \frac{\hbar}{2} \sqrt{\frac{1}{\ln\left(\frac{x}{x_c}\right) }}\;\frac{1}{t_{pl}}, \\
\Delta \, t_{min} &=& 2 \, \sqrt{\frac{\hbar}{c^5}\, G\, \ln\left(\frac{x}{x_c}\right)} - \frac{\alpha}{c} \, \hbar = 2 \, \sqrt{\ln\left(\frac{x}{x_c}\right)}\; t_{pl} - \frac{\alpha}{c} \, \hbar.  
\eea
$\Delta \, E_{max}$ is related to $t_{pl}$. The physical value of  $\Delta \, t_{min}$ requires that 
\bea
 \alpha_0 &<& 2 \, M_{pl}\, \sqrt{\frac{G}{\hbar\, c}\, \ln\left(\frac{x}{x_c}\right)}=2\frac{M_p\, c^2}{\hbar} \, \sqrt{ \ln\left(\frac{x}{x_c}\right)}\; t_{pl}.
\eea
We notice that both maximal energy and minimal time are positive. Their dependence on $t_{pl}$ is very obvious. The minimal measurable time interval depends in $\alpha$, the parameter charactering the utilized GUP approach.

\subsection{At distances larger than $x_c$}

When the photon travels a distance $x$ larger than $x_c$, then the total uncertainty in time  is estimated as
\bea
\Delta \, t_{total} & \geq & \frac{\hbar}{2 \Delta E} + G \frac{\Delta E}{c^5} + 2 G  \frac{\Delta E}{c^5}  \ln\left(\frac{x}{x_c}\right).
\eea
The maximal measurable energy and corresponding minimal measurable time interval are given as
\bea
\Delta E_{max} &=& \sqrt{\frac{c^5 \hbar}{2 G [1+2 \ln(\frac{x}{x_c})] }} =  \sqrt{\frac{1}{2 [1+2 \ln(\frac{x}{x_c})] }}\; \hbar\; t_{pl}, \\
\Delta t_{min} &=& \sqrt{\frac{\hbar G (1+2 \ln(\frac{x}{x_c})) }{2 c^5}} + \sqrt{\frac{\hbar G}{2 c^5 (1+2 \ln(\frac{x}{x_c}))}} \left[\left(1+2 \ln\left(\frac{x}{x_c}\right)\right)\right] \nonumber \\
&=& \sqrt{\frac{2 \hbar G}{c^5} \left[1+2 \ln\left(\frac{x}{x_c}\right)\right]} = \sqrt{2 \left[1+2 \ln\left(\frac{x}{x_c}\right)\right]}\; t_{pl}.
\eea
The resulting $\Delta E_{min}$ and $\Delta t_{min}$ are finite and positive. Both quantities are related to $t_{pl}$. 

In quadratic GUP approach, the total uncertainty in time is estimated as
 \bea
 \Delta \, t_{total} & > & \frac{\hbar}{2\, \Delta E} - \frac{\alpha}{c}\, \hbar + 2\, \left(\frac{\alpha}{c}\right)^2 \, \hbar \, \Delta \, E + 2\, G \, \frac{\Delta \, E}{c^5} \ln\left(\frac{x}{x_c}\right) .
\eea
Accordingly, the maximal measurable energy and the related minimal measurable time interval interval are 
\bea
\Delta \, E_{max} &=& \frac{1}{2}\, \sqrt{A}, \\
\Delta \, t_{min} &=& 2 \frac{\hbar}{\sqrt{A}} - \frac{\alpha}{c} \, \hbar ,
\eea
where 
\bea
A &=& \frac{\hbar\, c^5}{G \ln\left(\frac{x}{x_c}\right)+\alpha^2\, c^3 \, \hbar} = \frac{\hbar^2}{\ln\left(\frac{x}{x_c}\right)+\alpha^2\, c^3 \, \hbar}\; \frac{1}{t_{pl}^2}. \label{eq:A}
\eea
The physical value of $\Delta t_{min}$ requires that
\bea 
\alpha &<& \frac{2\, c}{\sqrt{A}}, \label{eq:alfgxc1}
\eea
resulting in a fourth-order equation in $\alpha$. For instance, its solution would read
\bea
\alpha &<& M_p\, c\, \sqrt{\frac{\sqrt{16 c^3 \hbar + G^2 \left(\ln(x/c_c)\right)^2}}{2 c^2 \hbar}-\frac{G \ln(x/x_c)}{2 c^2 \hbar}},
\eea
or for simplicity 
\bea
\alpha_0 &<&  16\, c^4\, M_{p}\, \hbar.
\eea 
Although the photon is assumed to travel distances larger than $x_c$, we notice that the maximal measurable energy and minimal measurable time depend on $\alpha$. Also, we notice that $\alpha$, the parameter that characterizes the GUP approach, should remain finite at this large scale.

\section{Discussion}
\label{sec:disc}

Itzhaki considered the standard uncertainty principle and utilized the Schwarzshild solution in large scale in order to estimate the minimal measurable time interval \cite{itzhaki}. It was concluded that the uncertainty in time measurement depends on the distance separating the observer from the event, the clock's accuracy and size, and the time taken by photon to reach the observer. Introducing a distances, in which GR offers a good approach for QG, the minimum error in the time measurement was estimated. In the present work, we give estimations for the measurable maximum energy and the minimum of time interval due to GUP for a Schwarzschild black hole. First, we distinguish between linear and quadratic quadratic GUP approaches. Then, both approaches are implemented in estimating maximum energy and minimum time interval at two scales; QG and GR.

\subsection{Physical Insights of Higher Order GUP }

For a recent review on the experimental, physical and mathematical insights of the higher order GUP, the readers are advised to consult Ref. \cite{TD2013}. The existence of a minimal length and a maximum momentum (energy) accuracy is preferred by various physical observations \cite{TD2013}. In light of this, we recall that the GUP approach \cite{advplb} was originally assumed to fit well with the string theory and the black hole physics (with a quadratic term of momenta) and fits as well with the Doubly Special Relativity (DSR) (with a linear term of momenta). Furthermore, this approach seems to simultaneously predict the minimal measurable length and the maximum measurable momentum (energy) and suggest that the space should be quantized and/or discritized. A new GUP approach \cite{pedram} is characterized by a minimal length uncertainty and a maximal momentum (energy). Another approach is conjectured to absolve an extensive comparison with Kempf, Mangano and Mann (KMM) \cite{16}, which has been performed in Hilbert space \cite{Nouicer}. Here, a novel idea of minimal length modelled in terms of the quantized space-time was implemented. Thus, this new approach agrees well with the quantum field theory and Heisenberg algebra, especially in context of non-commutative coherent states representation. The resulting GUP approach can be studied at ultra-violet finiteness of Feynman propagator \cite{Nouicer}.

We give two examples on the physical implications. First, the GUP effects on the area law of the entropy have been analyzed \cite{Ali:2013ma}. This leads to a $\sqrt{\text{Area}}$-type correction to the area law of entropy which imply that the number of bits $N$ is modified. Therefore, a modification in Newton's law of gravitation was reported \cite{Ali:2013ma}. Surprisingly, this modification agrees with a different sign with the prediction of Randall-Sundrum II model which contains one uncompactified extra dimension. Such a modification may have observable consequences at length scales much larger than the Planck scale or even may contribute to explain MOND \cite{mond1}. 

Second, as GUP is based on a momentum-dependent modification in the standard dispersion relation, it would be conjectured to violate the principle of Lorentz invariance, as well.  From the resulting Hamiltonian, the velocity and time of flight of relativistic distant particles at Planck energy were derived \cite{Tawfik:2012hz}. Furthermore, It was discussed how QUP could potentially lead to observable experimental effects related to the violation of Lorentz invariance.

\subsection{Shortest Distance, General Relativity, Classical and Quantum Gravity}

In 1936, a novel idea that the gravity might not be a fundamental force was presented  by Bronstein \cite{Gorelik}. At that time, both weak and strong forces were not discovered. That the gravity does not allow an arbitrarily high concentration of mass in a small region of space-time makes it fundamentally different than the electrodynamics. Apparently, the concentration of mass in a small region of space-time leads to Schwarzshild singularity \cite{Gorelik}. The gravitational radius of the test-body $G\, ρ\,V/c^2$ used for measuring the minimal distance should by no means be larger than its linear dimensions $V^{1/3}$ \cite{Bronstein}. Thus, one obtains an upper bound for its density $\rho\, \lesssim c^{2}/G\, V^{2/3}$. Therefore, in this region the possibilities for measurements are even more restricted than one would conclude from the quantum-mechanical commutation relations \cite{cr25,cr27}. Without a profound change of the classical notions, it therefore seems hardly possible to extend the quantum theory of gravitation to this region. Bronstein wrote \cite{Bronstein,Sabine}: {\it The existence of quantum uncertainties in gravitational field is a strong argument for the necessity of quantizing it. It is very likely that a quantum theory of gravitation would then generalize these uncertainty relations to all other Christoffel symbols.} In 1960, the uncertainties in measuring the average values of Christoffel symbols due to the impossibility of concentrating a mass to a region smaller than its Schwarzschild radius were studied \cite{mead20}. Accordingly,  the conclusion of Bronstein was approved  \cite{Bronstein}.  In 1964, Mead realized the peculiar role of gravity to test physics at short distances \cite{maad,mead22}. He even showed that the role of gravity should not mean increasing in the Heisenberg measurement uncertainty. 

In QG, the spacetime participates in the interactions and even acquires quantum fluctuations. Despite the great success of GR, finding theory of QG that reconciles the continuous nature of gravitational fields with the inherent QM remains a challenge. Assuming Lorentz Invariance violation (LIV), Horava-Lifshitz gravity \cite{horava09} was suggested as a model for QG that could be experimentally tested. At high energy, the space and time are not anisotropic.  Compared to other approaches, such as Loop QG, Horava-Lifshitz gravity implements quantum critical phenomena, which can be studied in condensed matter physics \cite{solutHarava}. 

Loop QG is an approach in which curved spacetime is given as a grid of discrete (quantized) loops of gravitational field lines. At the Planck length scale, spacetime undergoes a spontaneous dimensional reduction to 2D (as a flat manifold) \cite{Carlip} and at a larger scale evolves back 4D, where GR describes it, well.

Itzhaki introduced $x_c$ \cite{itzhaki} to define the scale up to which GR remains a good approximation to QG. In the present work, we define an upper bound to $\alpha$ parameter, Eqs. (\ref{eq:alfgxc1}) and (\ref{eq:A}) up to which the quadratic GUP approach is physically applicable. The upper bound is given in term of  $x_c$. Increasing the scale from $x_c$ to $x$ is accompanied by a decrease in the minimum time 
\bea
\sqrt{2}\, t_{pl} &\longrightarrow & 2 \sqrt{\ln\left(\frac{x}{x_c}\right)}\; t_{pl} - \frac{\alpha}{c}\, \hbar, \label{eq:modT}
\eea 
implies that 
\bea
t_{pl} &=& -\frac{\alpha \, \hbar}{c \left[\sqrt{2} - 2 \sqrt{\ln\left(\frac{x}{x_c}\right)}\right]}, \\
\eea
where
$x/x_c > \exp(1/2)$.

\subsection{Consequences on the Physics in the Early Universe}

Recently, the effects of GUP on the inflationary dynamics and the thermodynamics of the early Universe have been studied \cite{Tawfik:2012he}. Accordingly, the tensorial and scalar density fluctuations in the inflation era are evaluated and compared to the standard case. A good agreement with the Wilkinson Microwave Anisotropy Probe data was reported \cite{Tawfik:2012he}. Assuming that a quantum gas of scalar particles is confined within a thin layer near the apparent horizon of the Friedmann-Lemaitre-Robertson-Walker (FLRW) Universe which satisfies the boundary condition, the number and entropy densities and the free energy arising form the quantum states are calculated using the GUP approach. A qualitative estimation for effects of the quantum gravity on all these thermodynamic quantities is introduced.

Assuming that a quantum gas of scalar particles is confined within a thin layer near the apparent horizon of the FLRW universe which satisfies the boundary condition, the number and entropy densities and the free energy arising form the quantum states are calculated using the GUP approach \cite{Tawfik:2012he}. Furthermore, a qualitative estimation for the effects of quantum gravity on all these thermodynamic quantities was introduced.

\subsection{Minimum Length and Quantized Mass}

The compact relativistic astrophysical objects and cosmological constant would be able to characterize the maximum mass and energy, respectively \cite{harko2005,harko2007}. Maximum mass and radius was found by Chandrasekhar and Landau \cite{chandra}; \hbox{$M_{max}\simeq [(\hbar c/G) m_B^{-4/3}]^{3/2}$} and \hbox{$R_{max}\leq (\hbar/m c)(\hbar c/G m_B^{2})^{1/2}$,} where $m_B$ ($m$) being the mass of baryon (electron or neutron). Six decades ago, Buchdahl estimated an absolute limit of the ratio mass-to-radius of a stable compact object \cite{buchdahl}; \hbox{$(M/R)(2G/c^2)\leq 8/9$}. On the other hand, the minimum mass $M$ and radius $R$ would be related to Plank mass \hbox{$m_{pl}=(\hbar c/ 2 G)^{1/2}$} and length \hbox{$l_{pl}=(\hbar G/ c^3)^{1/2}$} \cite{harko2005}. Two mass limits have been proposed \cite{wisson}:
\bea
m_P &\equiv & \left(\frac{h}{c}\right)\, \left(\frac{\Lambda}{3}\right)^{1/2} \;\;\; \text{is relevant to quantum scale and} \nn\\
m_E &\equiv & \left(\frac{c^2}{G}\right)\, \left(\frac{3}{\Lambda}\right)^{1/2}  \;\; \text{is relevant to cosmological scale},\nn
\eea
where $\Lambda$ is the cosmological constant. 
The latter is nothing but the mass of the observable Universe \cite{planck2013}. According to Wesson, the earlier gives the minimum mass or mass quanta, where quantized mass $m=(n \hbar/c) (\Lambda/3)^{1/2}$. Bohmer and Harko formulated a rigorous proof of a minimum mass (or density) in GR at finite positive cosmological constant  \cite{harko2005}
\bea
M &\geq & \frac{\Lambda\, c^2}{12\, G} \, R^3, \nn \\
\rho &\geq & \frac{\Lambda\, c^2}{16\, \pi\, G}.
\eea
The astrophysical consequences of these lower bounds include no compact object with mass and density lower than these values would exist and positive vacuum or dark energy (cosmological constant) is necessary.

In the present work, we show that the existence of a minimum length (related to minimum time) for instance could be a purely quantum effect. As given in Ref. \cite{harko2005,harko2007}, this was related to the existence of positive cosmological constant.

\section{Conclusions}
\label{sec:conc}

The Planck time is the constant $(G\,\hbar/c^5)^{1/2}$ with dimensions of {\it ''time''} formed from the gravitational constant $G$. We review previous attempts to estimate minimum measurable time and introduced the usage of GUP approaches in determining maximum energy and minimum time interval. The main conclusion is that the maximal measurable energy $\Delta E$ and minimal measurable time $\Delta t$ are related to $t_{pl}$ and therefore both are accordingly bounded. Itzhaki model uses the most simple time measurement process. It was concluded that any particles that will be added must necessarily increase the uncertainty of the metric without decreasing the minimal measurable time. Furthermore, Itzhaki summarized that measured uncertainty would represent a basic property of nature. 

In the present work, the possibility of finding measurable maximal energy and minimal time are estimated in different quantum aspects. First, we find that the linear generalized uncertainty principle (GUP) approach gives non-physical results. The resulting maximal energy $\Delta\, E$ violates the conservation of energy. The minimal time interval $\Delta\, t$ shows that the direction of the arrow of time is backward. In light of this, we conclude that the applicability of the linear GUP approach is accordingly limited or even altered. 

Second, we find that the quadratic GUP approach results in finite $\Delta\, E$ and positive $\Delta\, t$. Thus, this is utilized in calculating the maximal energy and minimal  time based on the Schwarzshild solution in large scale. The calculations are performed at the shortest distance for which the general relativity is assumed to be a good approximation for the quantum gravity and at  larger distances as well. It is found that both quantities are related to the Planck time. Then, the uncertainties in both quantities are accordingly bounded.

\section*{Acknowledgement}
The authors would like to thank the anonymous referees for their comments, suggestions and even critics, which considerably improved an earlier version of the manuscript! The work of AT was supported by the World Laboratory for Cosmology And Particle Physics (WLCAPP) http://wlcapp.net/. 
  
%%%%%%%%%%%%%%%%%%%%%%%%%%%%%%%%%%%%%%%%%%%%%%%%%%%%%%%%%%%%%%%%%%%%%%
%%%   References
%%%%%%%%%%%%%%%%%%%%%%%%%%%%%%%%%%%%%%%%%%%%%%%%%%%%%%%%%%%%%%%%%%%%%%

%-----------------------------------------------

%----------------------==================-------------------------

\end{document}